\documentclass[epj]{svjour}
\usepackage{graphics}
\usepackage{graphicx}
\usepackage{textcomp}
\usepackage{makeidx}
\usepackage{amsmath}
\usepackage{subfigure}
\usepackage{amssymb}
\usepackage{hyperref}
\usepackage{graphicx}
\usepackage{color}
\usepackage{float}
\usepackage{cite}
\usepackage{epstopdf}
\begin{document}
\title{A note of the first law of thermodynamics by gravitational decoupling}
\titlerunning{A note of the first law of thermodynamics by gravitational decoupling}
\author{Milko Estrada\inst{1}  \thanks{\emph{e-mail:} milko.estrada@gmail.com} \and Reginaldo Prado \inst{1} \thanks{\emph{e-mail:} reginaldo.prado@ua.cl}}
\authorrunning{M Estrada \and R Prado}
%

%
\institute{Departamento de F\'isica, Facultad de ciencias b\'asicas, Universidad de Antofagasta, Casilla 170, Antofagasta, Chile. }

\date{Received: date / Revised version: date}
%
\abstract{ We provide a way of decoupling the first law of thermodynamics in two sectors : the standard first law of thermodynamics and the quasi first law of thermodynamics. It is showed that both sectors share the same thermodynamics volume and the same entropy. However, the total thermodynamics pressure, the total temperature and the total local energy correspond to a simple sum of the thermodynamics contributions of each sector. Furthermore, it is showed a simple example, where there is a phase transition due to the behavior of the temperature at the quasi sector. }  

\PACS{
      {PACS-key}{discribing text of that key}   \and
      {PACS-key}{discribing text of that key}
     } 
%

\maketitle
\section{Introduction}

Finding new solutions of physical interest to the Einstein field equations is not an easy task due to the highly nonlinear behavior of its equations. In this regard, in (2017) was proposed the {\it Gravitational Decoupling Method} \cite{Ovalle:2017fgl}, which represents an easy algorithm to decouple gravitational sources in General Relativity. This algorithm involves a {\it Minimal Geometric Deformation} (MGD) to the metric tensor together with a {\it decoupling of sources}. The method was described in reference \cite{Ovalle:2017wqi} as follows: ``given two gravitational sources: a source A and an extra source B, {\it standard Einstein’s equations} are first solved for A, and then a simpler set of {\it quasi-Einstein} equations are solved for B. Finally, the two solutions can be combined in order to derive the complete solution for the {\it total system}". The source A represents to the {\it seed sector} and the source B represents to the {\it extra sector}. Since its appearance, this method has served to find several new solution of physical interest, as for example well behaved solutions that could represent stellar distributions \cite{Ovalle:2017wqi,Gabbanelli:2018bhs,Estrada:2018zbh,Heras:2018cpz,Sharif:2018toc,Morales:2018nmq,Morales:2018urp,Estrada:2018vrl,Maurya:2019wsk}, black hole solutions \cite{Ovalle:2018umz,Contreras:2018nfg,Contreras:2019iwm}, $f({\mathcal {G}})$ gravity \cite{Sharif:2018tiz}, Einstein Klein Gordon System \cite{Ovalle:2018ans}, Pure Lovelock gravity \cite{Estrada:2019aeh}, $f(R,T)$ gravity \cite{Maurya:2019hds}, $f(R)$ gravity \cite{Sharif:2019zan}. See other applications  \cite{Contreras:2018gzd,Panotopoulos:2018law,Contreras:2019fbk,Contreras:2019mhf,Heras:2019ibr,Gabbanelli:2019txr,Hensh:2019rtb,Leon:2019abq,Torres:2019mee,Rincon:2019jal,Casadio:2019usg,Sharif:2019mzv,Singh:2019ktp,Abellan:2020wjw}

On the other hand, the fact that the black holes, due to quantum fluctuations, emit as black bodies where its temperature is related to its surface gravity \cite{Hawking:1974sw,Bekenstein:1972tm,Bekenstein:1973ur,Hawking:1974rv}, shows that in these objects the geometry and thermodynamics are directly connected. In this regard, the simplest version of the first law of thermodynamics is (see for example \cite{Wald:1984rg}):
\begin{equation}\label{1aleySinPdV}
dM=TdS+\Omega dJ+ \phi dQ, 
\end{equation}
where $M$, $T$, $S$, $J$ and $Q$ represent the mass, temperature, entropy, angular momentum and electric charge, respectively. In equation \eqref{1aleySinPdV} one can notice the absence of the terms of pressure and volume. In that respect, one way to address this problem was showed in reference \cite{Kastor:2009wy}, where the cosmological constant was associated with the thermodynamics pressure of the system, $P=-\frac{\Lambda}{8\pi}$, whereas the thermodynamics volume corresponds to the thermodynamic variable
conjugate to $P$. So, the mass parameter is interpreted as the enthalpy. However, other way to address this problem was showed in reference \cite{Kothawala:2007em}, identifying the thermodynamics pressure with the radial pressure of the energy momentum tensor. As indicates this reference, it is showed an interesting `` analogy between the gravitational dynamics of the horizons and thermodynamics, specifically, it is showed that it is possible to interpret the field equations near any spherically symmetric horizon as a thermodynamic identity $dU=TdS-PdV$" for the non rotating and non charged case. In this way, the term $-PdV$ can represent the macroscopic work done by the system. This analogy has been used for example for the study of the evolution of the regular black holes with a cosmological horizon in theories of Lovelock with a unique ground state \cite{Aros:2019auf}, the study of regular black hole thermodynamics in Einstein Hilbert theory \cite{Dymnikova:2010zz}, stationary axis-symmetric
space time and time dependent evolving horizons \cite{Kothawala:2007em}, generic Lovelock theory \cite{Kothawala:2009kc} . See also \cite{Sheykhi:2010zz,Padmanabhan:2002sha,Cai:2007bh}. 

In this work, inspired by the gravitational decoupling method, where the equations of motion are characterized by the standard Einstein’s equations and the quasi Einstein equations, we write the equations of motion as a thermodynamics identity $dU=TdS-PdV$, which, in analogous way to the mentioned method, will be characterized by two sectors, namely, the {\it standard first law} and the {\it quasi first law}. We will determine, as consequence of the application of the algorithm, which thermodynamics quantities are shared for both sectors and which quantities of the total system correspond to a simple sum of the quantities of each sector. Furthermore, by mean of an example, where the energy density of the seed sector correspond to the Hayward model \cite{Hayward:2005gi} and the energy density of the extra sector correspond to the Dymnikova model \cite{Dymnikova:1992ux}, we will test the thermodynamics behavior of each sector and the total system.

\section{The Einstein Field Equations for multiple sources}
In this section, we will show the Einstein field equations for multiple sources. The seed energy momentum tensor $\bar{T}_{\mu \nu}$ is deformed by an additional source $\theta_{\mu \nu}$,  whose coupling is proportional to the constant $\alpha$ and causes anisotropic effects on the self gravitating system \cite{Estrada:2018vrl}. As indicates reference \cite{Ovalle:2017wqi}, this additional source can contain new fields, like scalar, vector and tensor fields. So, the energy momentum is:

\begin{equation} \label{TensorEnergiaMomentum}
    T_{\mu \nu} = \bar{T}_{\mu \nu} + \alpha \theta_{\mu \nu}, 
\end{equation}
where the conservation law is:
\begin{equation} \label{conservacion}
    \nabla_\nu T^{\mu \nu} =0.
\end{equation}

Taken account an energy momentum tensor of the form  $T^\mu_\nu=\mbox{diag}(-\rho,p_r,p_\theta,p_\theta,...)$ an a seed energy momentum tensor of the form $\bar{T}^\mu_\nu= \mbox{diag}(-\bar{\rho},\bar{p}_r,\bar{p}_\theta,\bar{p}_\theta,...)$, from equation \eqref{TensorEnergiaMomentum} it is straightforward to note that: 
\begin{align}
    \rho&=\bar{\rho}-\alpha \theta^0_0 \label{densidadeTotal} \\
    p_r &= \bar{p}_r +\alpha \theta^1_1 \label{presionradialTotal} \\
    p_\theta &= \bar{p}_{\theta} +\alpha \theta^2_2 \label{presiontangencialTotal}
\end{align}

The Einstein field equations, in natural units, correspond to
\begin{equation}
    G^\mu_\nu = 8 \pi T^\mu_\nu.
\end{equation}

We will study the following spherically symmetric space time:
\begin{equation} \label{ElementoDeLinea}
    ds^2=-\mu(r)dt^2+\mu(r)^{-1} dr^2+r^2 d\Omega^2_{2}. 
\end{equation}

It is worth to mention that this metric satisfies the Einstein field equations under the following constraint to the energy momentum tensor:
\begin{align}
  T^t_t& =T^r_r  \\
  T^\theta_\theta&=T^\phi_\phi,
\end{align}
which implies that 
\begin{align}
    \rho&=-p_r  \label{condicionTotal1}\\
    p_\theta &=p_\phi. \label{CondicionTotal2}
\end{align}

To accomplish the condition \eqref{condicionTotal1} it is imposed in arbitrarily way that 
\begin{align} 
    \bar{\rho}&=-\bar{p}_r \label{CondicionParcial1} \mbox{   ,and} \\
       \theta^0_0&=\theta^1_1 \label{CondicionParcial2}, 
\end{align}
on the other hand , the condition $T^\theta_\theta=T^\phi_\phi$ arises from spherical symmetry.

In this way, the $(t-t)$ and $(r-r)$ components of the Einstein field equations are similar and are given by:
\begin{equation} \label{EcuacionDeMovimiento1}
   8\pi \bar{\rho}-8 \pi \alpha \theta^0_0 = \dfrac{1-\mu}{r^2}-\dfrac{\mu'}{r},
\end{equation}
where $'$ denotes derivation respect to the radial coordinate.

On the other hand, the conservation law has the form:
\begin{align} \label{Conservacion1}
&\bar{p}_r'+ \frac{2}{r}(\bar{p}_r-\bar{p}_t ) \nonumber \\
&+ \alpha \Big ( (\theta^1_1)'+\frac{2}{r}(\theta^1_1-\theta^2_2 ) \Big )=0 .
\end{align}
As indicates the reference \cite{Ovalle:2017fgl}, due that, in the limit $\alpha \to 0$, the geometry of the line element \eqref{ElementoDeLinea} is associate with the geometry of the seed perfect fluid, and its respective Bianchi identities are satisfied, the seed energy momentum tensor is conserved, {\it i.e.} :
$\nabla_\nu \bar{T}^{\mu \nu} =0$:
\begin{equation} \label{Conservacion2}
\bar{p}_r'+ \frac{2}{r}(\bar{p}_r-\bar{p}_t ) =0  ,
\end{equation}
and, by inserting equation \eqref{Conservacion2} into equation \eqref{Conservacion1} , it is easily to see that the extra source is also conserved, {\it i.e.} $\nabla_\nu \theta^{\mu \nu} =0$:
\begin{equation}
     (\theta^1_1)'+\frac{2}{r}(\theta^1_1-\theta^2_2 ) =0,
\end{equation}
thus, each source is separately conserved.

\section{Decoupling the first law of thermodynamics}

In order to analyse the thermodynamics, we will call $r=a$ to the generic horizon, where $\mu(a)=0$. The first step, following reference \cite{Kothawala:2007em}, is, identify the thermodynamics pressure with the radial pressure of the fluid perfect, 
\begin{align}
    P_{tot}=p_r=&\bar{p}_r + \alpha \theta^1_1 \nonumber \\
        P_{tot}=&P_s+ \alpha P_q,
\end{align} 
where $\bar{p}_r=P_s$ and $\theta^0_0=P_q$, and, using the conditions \eqref{CondicionParcial1} and \eqref{CondicionParcial2},  evaluating the equations of motion \eqref{EcuacionDeMovimiento1} at $r=a$:
\begin{equation} \label{EcuacionTermo1}
    4\pi a^2 \left ( P_s + \alpha P_q \right )= \dfrac{u'}{2}a- \frac{1}{2}.
\end{equation}

Next, also following \cite{Kothawala:2007em}, we consider two horizons whose values are $a$ and $a+da$, thus, multiplying equation \eqref{EcuacionTermo1} by $da$: 
\begin{equation} \label{EcuacionTermo2}
    (P_s+ \alpha P_q) d \left ( \frac{4}{3}\pi a^3   \right) = \left (  \dfrac{\mu'}{4\pi} \right) d \left ( \dfrac{4\pi a^2}{4} \right ) - d \left ( \dfrac{a}{2}  \right) . 
\end{equation}

To test the role of the constant $\alpha$, we will apply the MGD to the geometry. Turning on $\alpha$, the effects of the source $\theta_{\mu \nu}$ appear on the seed solution. These effects are encoded in the geometric deformation undergone by the seed fluid geometry \eqref{ElementoDeLinea} as follows:
\begin{equation} \label{Deformacion}
    \mu (r) \to  \bar{\mu} (r) + \alpha g(r). 
\end{equation}

Thus, it is direct to see that the equation \eqref{EcuacionDeMovimiento1} can be written as:
\begin{equation} \label{EcuacionDeMovimiento2}
   \mu= \bar{\mu} + \alpha g =1 - \dfrac{2}{r} \Big ( m_s(r) + \alpha m_q(r)    \Big ),
\end{equation}
where
\begin{align}
    m_s(r)=& \int 4 \pi r^2 \bar{\rho} dr, \\
    m_q(r)=&-\int 4 \pi r^2 \theta^0_0 dr ,
\end{align}
and
\begin{align}
    \bar{\mu}(r)=&1 - \dfrac{2}{r}m_s(r) \\
           g(r)=& - \dfrac{2}{r}m_q(r).
\end{align}

To ensure an asymptotically flat behavior, must be imposed that:
\begin{align}
    \displaystyle & \lim_{r \to \infty}  m_s(r) =M_s \label{CondicionInfinito1} \\
    \displaystyle & \lim_{r \to \infty}  m_q(r) =M_q \label{CondicionInfinito2}, 
\end{align}
also
\begin{align}
    \displaystyle & \lim_{r \to \infty}  \frac{d}{dr} m_s(r) =0 \\
    \displaystyle & \lim_{r \to \infty}  \frac{d}{dr} m_q(r) =0 ,
\end{align}

Thus, replacing \eqref{EcuacionDeMovimiento2} at $r=a$, where $\mu(a)=0$:
\begin{equation} \label{amedios}
    \dfrac{a}{2}= m_s(a) + \alpha \cdot m_q(a).
\end{equation}

Thus, inserting equations \eqref{Deformacion} and \eqref{amedios} into equation \eqref{EcuacionTermo2}, this last equation takes the following form:
\begin{align}
    (P_s+ \alpha P_q) d \left ( \frac{4}{3}\pi a^3   \right) =& \left (  \dfrac{\bar{\mu}'}{4\pi}+ \alpha \dfrac{g'}{4\pi} \right) d \left ( \dfrac{4\pi a^2}{4} \right ) \nonumber \\
    &- d \Big ( m_s(a) + \alpha \cdot m_q(a)  \Big ) . \end{align}

Thus, the system splits into the following sets of equations:
\begin{itemize}
    \item The {\it standard firs law of thermodynamics} of order $\alpha^0$, which corresponds to the standard Einstein equations:
    \begin{equation}
        P_s d \left ( \frac{4}{3}\pi a^3   \right) = \left (  \dfrac{\bar{\mu}'}{4\pi} \right) d \left ( \dfrac{4\pi a^2}{4} \right )- d \Big ( m_s(a) \Big ).
    \end{equation}

This equation has the form of the first law of thermodynamics $PdV=TdS-dU$, where each term is identified as:
\begin{align}
    \mbox{Thermodynamics Pressure}=&P_s \\
    \mbox{Volume}=& \frac{4}{3}\pi a^3 \\
    \mbox{Temperature}=& T_s=\frac{\bar{\mu}'}{4\pi} \\
    \mbox{Entropy}=& \dfrac{4\pi a^2}{4}=\dfrac{Area}{4} \\
    \mbox{Local Energy}=&U_s^{loc}= m_s(a)
    \end{align}

It is worth to mention that our thermodynamics variables obtained coincide with the previously known in literature. The thermodynamics volume coincide with the geometric volume, the temperature is the well known expression and the entropy follows the area's law. The  energy $m_s(a)$ corresponds to a local definition of energy, at $r=a$, given by reference \cite{Aros:2019auf}. In the vacuum case, where $\bar{\rho}=\theta^\mu_\nu=0$, the solution \eqref{EcuacionDeMovimiento2} has the form $\mu=1-2M/r$ and the local energy coincide with the energy of the Schwarzschild black hole, $a/2=M$. 

These thermodynamics variables $P_s$,$T_s$ and $U_s^{loc}$ represent the contribution of the standard sector to the total pressure, total temperature and total local energy of the black hole. These ones do not represent the pressure, temperature and local energy of an independent black hole.

\item The {\it quasi first law of thermodynamics} of order $\alpha$, which corresponds to the quasi Einstein equations:

 \begin{equation}
        P_q d \left ( \frac{4}{3}\pi a^3   \right) = \left (  \dfrac{g'}{4\pi} \right) d \left ( \dfrac{4\pi a^2}{4} \right )- d \Big ( m_q(a) \Big ).
    \end{equation}

This equation also has the form of the first law of thermodynamics,  $PdV=TdS-dU$, and each term is identified as:
\begin{align}
    \mbox{Thermodynamics Pressure}=&P_q \\
    \mbox{Volume}=& \frac{4}{3}\pi a^3 \\
    \mbox{Temperature}=& T_q=\frac{g'}{4\pi} \\
    \mbox{Entropy}=& \dfrac{4\pi a^2}{4}=\dfrac{Area}{4} \\
    \mbox{Local Energy}=&U_q^{loc}= m_q(a),
    \end{align}
where, in a similar way to the previous case, all the thermodynamics variables coincide with the previously known in the literature.

Also, these thermodynamics variables $P_q$,$T_q$ and $U_q^{loc}$ represent the contribution of the quasi sector to the total pressure, total temperature and total local energy of the black hole. These ones do not represent the pressure, temperature and local energy of an independent black hole.

\end{itemize}

\subsection{Some Remarks}
From the previous analysis one can remark some properties:
\begin{itemize}
    \item The total thermodynamics pressure, total temperature and total local energy correspond to a  simple sum of the thermodynamics contributions of each sector
    \begin{align}
        P_{tot}&=P_s+P_q \label{Ptotal} \\
        T_{tot}&=T_s+T_q \label{Ttotal}\\
        U_{tot}^{loc}&=U_s^{loc}+U_q^{loc} \label{Utotal}
    \end{align}

So, as was said above, it is worth to mention that the temperatures $T_s$ and $T_q$ represent the contribution of each sector (standard and quasi) to the total temperature of black hole and, do not represent the temperature of one independent black hole. The same occurs with the pressure and energy.

    \item Both sectors share the same thermodynamics volume and entropy.
\end{itemize}
As was said above, $U_{tot}^{loc}$ corresponds to a local definition of energy, located at $r=a$ \cite{Aros:2019auf}. However, the total energy of the system must be computed by the Noether charge \cite{Aros:1999kt}, by mean of the Komar formula \cite{Kastor:2008xb}.

\begin{align}
   E \propto & \displaystyle \lim_{r \to \infty} K(\xi)= \nonumber \\
   =&\lim_{r \to \infty} \frac{1}{8\pi} \frac{d}{dr} \mu(r) r^{2} \int d\Omega_{2} \nonumber \\
   =& \dfrac{1}{2} \Big (M_s+ \alpha \cdot  M_q \Big ),
\end{align}
{\it i.e} the total energy is proportional to the Komar formula \cite{Kastor:2008xb}. After a regularization, based on the inclusion of the boundary terms in the action \cite{Aros:1999kt}, can be obtained that:
\begin{equation} \label{EcuacionEnergia}
    E= M_s+ \alpha \cdot  M_q. 
\end{equation}

At infinity, our local definitions of energy coincide with the total energy. Thus, other important result of our decoupling of the first law of thermodynamics is that: The total energy corresponds to the contribution of the Noether charge of each sector.

\section{A simple example}
As a simple example, we choose a seed and a extra source, such that, near the origin, the mass functions behave as:
\begin{align}
   & m_s(r) \Big |_{r \approx 0} \approx C_1 r^3 \label{CondicionRegular1} \\
   & m_q(r) \Big |_{r \approx 0} \approx C_2 r^3. \label{CondicionRegular2}
\end{align}

This one implies that near the origin the geometry behaves as a de Sitter space time, where the invariants of curvature have finite values, and thus, unlike the singular Schwarzschild solution, the geometry is regular near the origin. These models are called {\it regular black holes}. 

The seed source is given by the Hayward model \cite{Hayward:2005gi}:
\begin{equation}
    \rho(r) = \frac{1}{2\pi} \frac{3Q^2M_s^2}{(2Q^2M_s+r^3)^2},
\end{equation}
where $Q$ is a constant. The mass function is: 
\begin{equation}
    m(r)=\frac{M_sr^3}{r^3+2M_sQ^2}.
\end{equation}

On the other hand, the extra source is given by the Dymnikova model \cite{Dymnikova:2010zz}:
\begin{equation}
    \theta^0_0(r) = -\frac{M_q \exp(-r^3/R^3)}{(4/3)\pi R^3},
\end{equation}
where $R$ is a constant. The mass function is:
\begin{equation}
    m(r)=M_q \Big (1-\exp(-r^3/R^3) \Big ).
\end{equation}

It is direct to check that the conditions \ref{CondicionRegular1} and \ref{CondicionRegular2} are fulfilled, an thus, this model is suited to represent a regular black hole. 

On the other hand, also it is direct to check that the conditions \ref{CondicionInfinito1} and \ref{CondicionInfinito2} are fulfilled. This one ensures that the total energy corresponds to the contribution of the Noether charge of each sector, equation \eqref{EcuacionEnergia}.

\begin{figure}
\centering
{\includegraphics[width=3in]{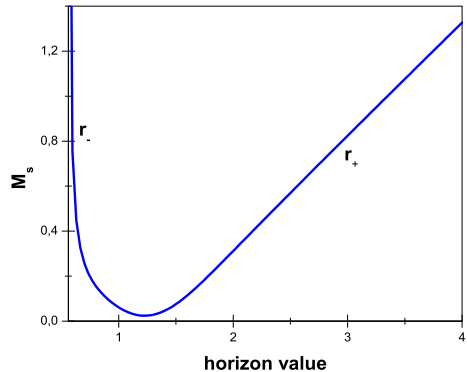}}
\caption{Behavior of $M_s$ parameter for $\alpha=Q=R1$, $M_q=0.7$}
\label{GraficoM}
\end{figure}

In the figure \ref{GraficoM} one can check the existence of one critical value of $M_s=M_*$, which corresponds to the minimum value on the curve, where there is an extremal black hole, where the internal horizon $a=r_-$ and black hole horizon $a=r_+$ coincide. On the other hand, for $M_s>M_*$ there are two horizons, namely, the internal horizon $r_-$ and the black hole horizon $r_+$. This behavior is generic for other values of the parameters.

\begin{figure}
\centering
\subfigure[$T_s$ .]{\includegraphics[width=75mm]{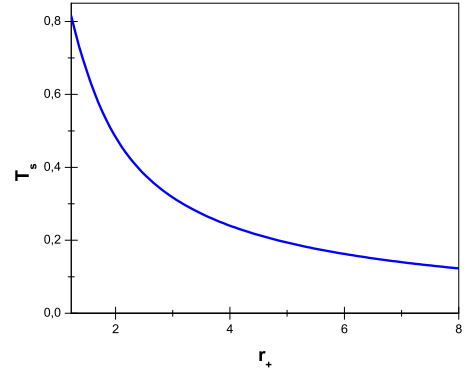}} 
\subfigure[$T_q$ ]{\includegraphics[width=75mm]{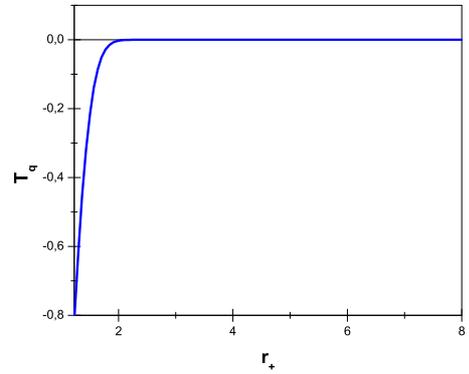}}
\subfigure[$T_{tot}$ ]{\includegraphics[width=75mm]{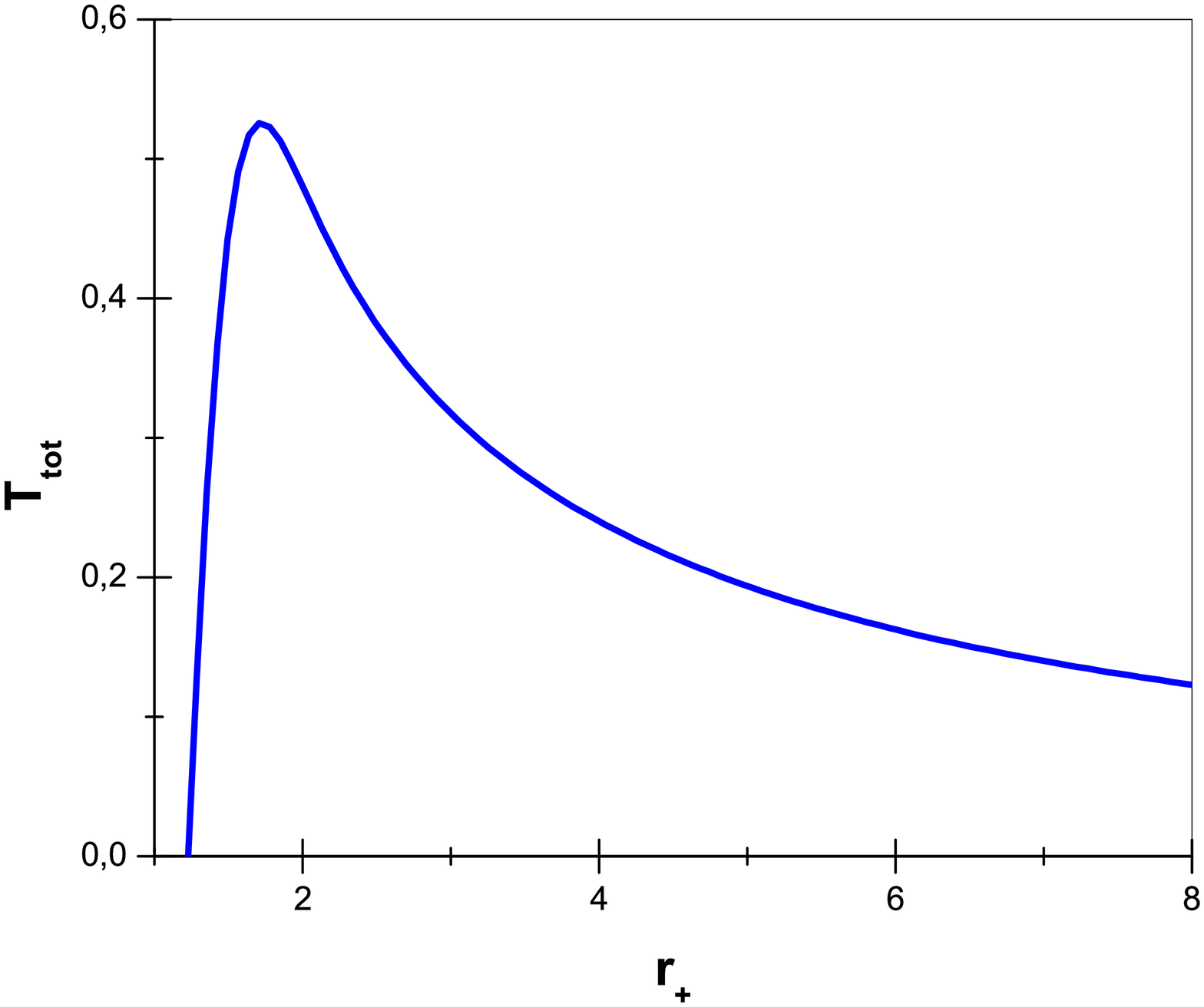}}
\caption{Behavior of $T_s$,$T_q$ and $T_{tot}$ for $\alpha=Q=R1$, $M_q=0.7$ .}
\label{FigTemperatura}
\end{figure}

Figures \ref{FigTemperatura} (a), \ref{FigTemperatura}(b) and \ref{FigTemperatura}(c) display the behavior of the contributions $T_s$ and $T_q$ and the total temperature $T_{tot}$, respectively. This behavior is generic for other values of the parameters. One can check that the temperature vanishes at the extremal case \cite{Chirco:2010rq}. $T_s$ and $T_q$ are contributions to the total temperature $T_{tot}$, where, this last is always positive. The total temperature has a local maximum, located at $r_+=r_{max}$. One can check that this inflection point is due to the behavior of the contribution of the quasi sector, $T_q$, which vanishes after this point. Using the standard definition of the heat capacity:
\begin{equation}
    C= \frac{\partial S}{\partial r_+} \left ( \frac{\partial T_{tot}}{\partial r_+}    \right )^{-1} ,
\end{equation}
it is direct to see that the sign of the specific heat depend only on the factor $\dfrac{\partial T_{tot}}{\partial r_+}$. So in the point $r_+=r_{max}$, where this derivative vanishes, there is a phase transition. At the left side of $r_+=r_{max}$ this derivative is positive and the heat capacity is positive, {\it i.e} the black hole is stable. However, at the right side of $r_+=r_{max}$ this derivative is negative and the heat capacity is negative, {\it i.e} the black hole is unstable. 

\section{Conclusion and Discussion}

In this work is showed that, a direct consequence of the application of the gravitational decoupling method to the spherically symmetric space time is the decoupling of the first law of thermodynamics in two sectors, called, the standard first law of thermodynamics, and the quasi first law of thermodynamics. This is achieved, following the approximation of the reference \cite{Kothawala:2007em}, where the equations of motion are evaluated at the horizon $r=a$. 

Each sector is written as a thermodynamics identity $PdV=TdS-dU$. In this respect, both sectors share the same thermodynamics volume and the same entropy. The total thermodynamics pressure, total temperature and total local energy correspond to a  simple sum of the thermodynamics contributions of each sector, equations \eqref{Ptotal}, \eqref{Ttotal} and \eqref{Utotal}. Other interesting result is that the total energy corresponds to the contribution of the Noether charge of each sector, equation \ref{EcuacionEnergia}.

In both sectors, the terms corresponding to the identity $PdV=TdS-dU $ coincide with the previously known in literature. The thermodynamics volume coincide with the geometric volume, the temperature is the well known expression and the entropy follows the area's law. The local definition of energy coincide with the definition of the reference \cite{Aros:2019auf}.

Finally, it is provided one example, where the seed source and the extra sources correspond to the Hayward model \cite{Hayward:2005gi} and the Dymnikova model \cite{Dymnikova:1992ux}, respectively. It is showed that the total temperature has a inflection point where there is a phase transition between stable/unstable black hole. This phase transition is due to the behavior of the quasi sector contribution, where this contribution vanishes after this point.

\bibliography{mybib}

\begin{thebibliography}{10}
\expandafter\ifx\csname url\endcsname\relax
  \def\url#1{\texttt{#1}}\fi
\expandafter\ifx\csname urlprefix\endcsname\relax\def\urlprefix{URL }\fi
\expandafter\ifx\csname href\endcsname\relax
  \def\href#1#2{#2} \def\path#1{#1}\fi

\bibitem{Ovalle:2017fgl}
J.~Ovalle, {Decoupling gravitational sources in general relativity: from
  perfect to anisotropic fluids}, Phys. Rev. D95~(10) (2017) 104019.
\newblock \href {http://arxiv.org/abs/1704.05899} {\path{arXiv:1704.05899}},
  \href {http://dx.doi.org/10.1103/PhysRevD.95.104019}
  {\path{doi:10.1103/PhysRevD.95.104019}}.

\bibitem{Ovalle:2017wqi}
J.~Ovalle, R.~Casadio, R.~da~Rocha, A.~Sotomayor, {Anisotropic solutions by
  gravitational decoupling}, Eur. Phys. J. C78~(2) (2018) 122.
\newblock \href {http://arxiv.org/abs/1708.00407} {\path{arXiv:1708.00407}},
  \href {http://dx.doi.org/10.1140/epjc/s10052-018-5606-6}
  {\path{doi:10.1140/epjc/s10052-018-5606-6}}.

\bibitem{Gabbanelli:2018bhs}
L.~Gabbanelli, A.~Rincon, C.~Rubio, {Gravitational decoupled anisotropies in
  compact stars}, Eur. Phys. J. C78~(5) (2018) 370.
\newblock \href {http://arxiv.org/abs/1802.08000} {\path{arXiv:1802.08000}},
  \href {http://dx.doi.org/10.1140/epjc/s10052-018-5865-2}
  {\path{doi:10.1140/epjc/s10052-018-5865-2}}.

\bibitem{Estrada:2018zbh}
M.~Estrada, F.~Tello-Ortiz, {A new family of analytical anisotropic solutions
  by gravitational decoupling}, Eur. Phys. J. Plus 133~(11) (2018) 453.
\newblock \href {http://arxiv.org/abs/1803.02344} {\path{arXiv:1803.02344}},
  \href {http://dx.doi.org/10.1140/epjp/i2018-12249-9}
  {\path{doi:10.1140/epjp/i2018-12249-9}}.

\bibitem{Heras:2018cpz}
C.~L. Heras, P.~Leon, {Using MGD gravitational decoupling to extend the
  isotropic solutions of Einstein equations to the anisotropical domain},
  Fortsch. Phys. 66~(7) (2018) 1800036.
\newblock \href {http://arxiv.org/abs/1804.06874} {\path{arXiv:1804.06874}},
  \href {http://dx.doi.org/10.1002/prop.201800036}
  {\path{doi:10.1002/prop.201800036}}.

\bibitem{Sharif:2018toc}
M.~Sharif, S.~Sadiq, {Gravitational Decoupled Charged Anisotropic Spherical
  Solutions}, Eur. Phys. J. C78~(5) (2018) 410.
\newblock \href {http://arxiv.org/abs/1804.09616} {\path{arXiv:1804.09616}},
  \href {http://dx.doi.org/10.1140/epjc/s10052-018-5894-x}
  {\path{doi:10.1140/epjc/s10052-018-5894-x}}.

\bibitem{Morales:2018nmq}
E.~Morales, F.~Tello-Ortiz, {Charged anisotropic compact objects by
  gravitational decoupling}, Eur. Phys. J. C78~(8) (2018) 618.
\newblock \href {http://arxiv.org/abs/1805.00592} {\path{arXiv:1805.00592}},
  \href {http://dx.doi.org/10.1140/epjc/s10052-018-6102-8}
  {\path{doi:10.1140/epjc/s10052-018-6102-8}}.

\bibitem{Morales:2018urp}
E.~Morales, F.~Tello-Ortiz, {Compact Anisotropic Models in General Relativity
  by Gravitational Decoupling}, Eur. Phys. J. C78~(10) (2018) 841.
\newblock \href {http://arxiv.org/abs/1808.01699} {\path{arXiv:1808.01699}},
  \href {http://dx.doi.org/10.1140/epjc/s10052-018-6319-6}
  {\path{doi:10.1140/epjc/s10052-018-6319-6}}.

\bibitem{Estrada:2018vrl}
M.~Estrada, R.~Prado, {The Gravitational decoupling method: the higher
  dimensional case to find new analytic solutions}, Eur. Phys. J. Plus 134~(4)
  (2019) 168.
\newblock \href {http://arxiv.org/abs/1809.03591} {\path{arXiv:1809.03591}},
  \href {http://dx.doi.org/10.1140/epjp/i2019-12555-8}
  {\path{doi:10.1140/epjp/i2019-12555-8}}.

\bibitem{Maurya:2019wsk}
S.~K. Maurya, F.~Tello-Ortiz, {Generalized relativistic anisotropic compact
  star models by gravitational decoupling}, Eur. Phys. J. C79~(1) (2019) 85.
\newblock \href {http://dx.doi.org/10.1140/epjc/s10052-019-6602-1}
  {\path{doi:10.1140/epjc/s10052-019-6602-1}}.

\bibitem{Ovalle:2018umz}
J.~Ovalle, R.~Casadio, R.~d. Rocha, A.~Sotomayor, Z.~Stuchlik, {Black holes by
  gravitational decoupling}, Eur. Phys. J. C78~(11) (2018) 960.
\newblock \href {http://arxiv.org/abs/1804.03468} {\path{arXiv:1804.03468}},
  \href {http://dx.doi.org/10.1140/epjc/s10052-018-6450-4}
  {\path{doi:10.1140/epjc/s10052-018-6450-4}}.

\bibitem{Contreras:2018nfg}
E.~Contreras, P.~Bargueño, {Minimal Geometric Deformation in asymptotically
  (A-)dS space-times and the isotropic sector for a polytropic black hole},
  Eur. Phys. J. C78~(12) (2018) 985.
\newblock \href {http://arxiv.org/abs/1809.09820} {\path{arXiv:1809.09820}},
  \href {http://dx.doi.org/10.1140/epjc/s10052-018-6472-y}
  {\path{doi:10.1140/epjc/s10052-018-6472-y}}.

\bibitem{Contreras:2019iwm}
E.~Contreras, A.~Rincon, P.~Bargueño, {A general interior anisotropic solution
  for a BTZ vacuum in the context of the Minimal Geometric Deformation
  decoupling approach}, Eur. Phys. J. C79~(3) (2019) 216.
\newblock \href {http://arxiv.org/abs/1902.02033} {\path{arXiv:1902.02033}},
  \href {http://dx.doi.org/10.1140/epjc/s10052-019-6749-9}
  {\path{doi:10.1140/epjc/s10052-019-6749-9}}.

\bibitem{Sharif:2018tiz}
M.~Sharif, S.~Saba, {Gravitational decoupled anisotropic solutions in
  $f({\mathcal {G}})$ gravity}, Eur. Phys. J. C78~(11) (2018) 921.
\newblock \href {http://arxiv.org/abs/1811.08112} {\path{arXiv:1811.08112}},
  \href {http://dx.doi.org/10.1140/epjc/s10052-018-6406-8}
  {\path{doi:10.1140/epjc/s10052-018-6406-8}}.

\bibitem{Ovalle:2018ans}
J.~Ovalle, R.~Casadio, R.~da~Rocha, A.~Sotomayor, Z.~Stuchlik,
  {Einstein-Klein-Gordon system by gravitational decoupling}, EPL 124~(2)
  (2018) 20004.
\newblock \href {http://arxiv.org/abs/1811.08559} {\path{arXiv:1811.08559}},
  \href {http://dx.doi.org/10.1209/0295-5075/124/20004}
  {\path{doi:10.1209/0295-5075/124/20004}}.

\bibitem{Estrada:2019aeh}
M.~Estrada, {A way of decoupling gravitational sources in pure Lovelock
  gravity}, Eur. Phys. J. C79~(11) (2019) 918.
\newblock \href {http://arxiv.org/abs/1905.12129} {\path{arXiv:1905.12129}},
  \href {http://dx.doi.org/10.1140/epjc/s10052-019-7444-6}
  {\path{doi:10.1140/epjc/s10052-019-7444-6}}.

\bibitem{Maurya:2019hds}
S.~K. Maurya, F.~Tello-Ortiz, {Charged anisotropic compact star in $f(R,T)$
  gravity: A minimal geometric deformation gravitational decoupling approach},
  Phys. Dark Univ. 27 (2020) 100442.
\newblock \href {http://arxiv.org/abs/1905.13519} {\path{arXiv:1905.13519}},
  \href {http://dx.doi.org/10.1016/j.dark.2019.100442}
  {\path{doi:10.1016/j.dark.2019.100442}}.

\bibitem{Sharif:2019zan}
M.~Sharif, A.~Waseem, {Effects of Charge on Gravitational Decoupled Anisotropic
  Solutions in f(R) Gravity}, Chin. J. Phys. 60 (2019) 426--439.
\newblock \href {http://arxiv.org/abs/1906.07559} {\path{arXiv:1906.07559}},
  \href {http://dx.doi.org/10.1016/j.cjph.2019.05.016}
  {\path{doi:10.1016/j.cjph.2019.05.016}}.

\bibitem{Contreras:2018gzd}
E.~Contreras, {Minimal Geometric Deformation: the inverse problem}, Eur. Phys.
  J. C78~(8) (2018) 678.
\newblock \href {http://arxiv.org/abs/1807.03252} {\path{arXiv:1807.03252}},
  \href {http://dx.doi.org/10.1140/epjc/s10052-018-6168-3}
  {\path{doi:10.1140/epjc/s10052-018-6168-3}}.

\bibitem{Panotopoulos:2018law}
G.~Panotopoulos, A.~Rincon, {Minimal Geometric Deformation in a cloud of
  strings}, Eur. Phys. J. C78~(10) (2018) 851.
\newblock \href {http://arxiv.org/abs/1810.08830} {\path{arXiv:1810.08830}},
  \href {http://dx.doi.org/10.1140/epjc/s10052-018-6321-z}
  {\path{doi:10.1140/epjc/s10052-018-6321-z}}.

\bibitem{Contreras:2019fbk}
E.~Contreras, {Gravitational decoupling in $2+1$ dimensional space--times with
  cosmological term}, Class. Quant. Grav. 36~(9) (2019) 095004.
\newblock \href {http://arxiv.org/abs/1901.00231} {\path{arXiv:1901.00231}},
  \href {http://dx.doi.org/10.1088/1361-6382/ab11e6}
  {\path{doi:10.1088/1361-6382/ab11e6}}.

\bibitem{Contreras:2019mhf}
E.~Contreras, P.~Bargueño, {Extended gravitational decoupling in 2 + 1
  dimensional space-times}, Class. Quant. Grav. 36~(21) (2019) 215009.
\newblock \href {http://arxiv.org/abs/1902.09495} {\path{arXiv:1902.09495}},
  \href {http://dx.doi.org/10.1088/1361-6382/ab47e2}
  {\path{doi:10.1088/1361-6382/ab47e2}}.

\bibitem{Heras:2019ibr}
C.~Las~Heras, P.~León, {New algorithms to obtain analytical solutions of
  Einstein’s equations in isotropic coordinates}, Eur. Phys. J. C79~(12)
  (2019) 990.
\newblock \href {http://arxiv.org/abs/1905.02380} {\path{arXiv:1905.02380}},
  \href {http://dx.doi.org/10.1140/epjc/s10052-019-7507-8}
  {\path{doi:10.1140/epjc/s10052-019-7507-8}}.

\bibitem{Gabbanelli:2019txr}
L.~Gabbanelli, J.~Ovalle, A.~Sotomayor, Z.~Stuchlik, R.~Casadio, {A causal
  Schwarzschild-de Sitter interior solution by gravitational decoupling}, Eur.
  Phys. J. C79~(6) (2019) 486.
\newblock \href {http://arxiv.org/abs/1905.10162} {\path{arXiv:1905.10162}},
  \href {http://dx.doi.org/10.1140/epjc/s10052-019-7022-y}
  {\path{doi:10.1140/epjc/s10052-019-7022-y}}.

\bibitem{Hensh:2019rtb}
S.~Hensh, Z.~Stuchlík, {Anisotropic Tolman VII solution by gravitational
  decoupling}, Eur. Phys. J. C79~(10) (2019) 834.
\newblock \href {http://arxiv.org/abs/1906.08368} {\path{arXiv:1906.08368}},
  \href {http://dx.doi.org/10.1140/epjc/s10052-019-7360-9}
  {\path{doi:10.1140/epjc/s10052-019-7360-9}}.

\bibitem{Leon:2019abq}
P.~León, A.~Sotomayor, {Braneworld Gravity under gravitational decoupling},
  Fortsch. Phys. 67~(12) (2019) 1900077.
\newblock \href {http://arxiv.org/abs/1907.11763} {\path{arXiv:1907.11763}},
  \href {http://dx.doi.org/10.1002/prop.201900077}
  {\path{doi:10.1002/prop.201900077}}.

\bibitem{Torres:2019mee}
V.~A. Torres-Sánchez, E.~Contreras, {Anisotropic neutron stars by
  gravitational decoupling}, Eur. Phys. J. C79~(10) (2019) 829.
\newblock \href {http://arxiv.org/abs/1908.08194} {\path{arXiv:1908.08194}},
  \href {http://dx.doi.org/10.1140/epjc/s10052-019-7341-z}
  {\path{doi:10.1140/epjc/s10052-019-7341-z}}.

\bibitem{Rincon:2019jal}
A.~Rincon, L.~Gabbanelli, E.~Contreras, F.~Tello-Ortiz, {Minimal geometric
  deformation in a Reissner–Nordström background}, Eur. Phys. J. C79~(10)
  (2019) 873.
\newblock \href {http://arxiv.org/abs/1909.00500} {\path{arXiv:1909.00500}},
  \href {http://dx.doi.org/10.1140/epjc/s10052-019-7397-9}
  {\path{doi:10.1140/epjc/s10052-019-7397-9}}.

\bibitem{Casadio:2019usg}
R.~Casadio, E.~Contreras, J.~Ovalle, A.~Sotomayor, Z.~Stuchlick,
  {Isotropization and change of complexity by gravitational decoupling}, Eur.
  Phys. J. C79~(10) (2019) 826.
\newblock \href {http://arxiv.org/abs/1909.01902} {\path{arXiv:1909.01902}},
  \href {http://dx.doi.org/10.1140/epjc/s10052-019-7358-3}
  {\path{doi:10.1140/epjc/s10052-019-7358-3}}.

\bibitem{Sharif:2019mzv}
M.~Sharif, S.~Sadiq, {2+1-dimensional gravitational decoupled anisotropic
  solutions}, Chin. J. Phys. 60 (2019) 279--289.
\newblock \href {http://dx.doi.org/10.1016/j.cjph.2019.05.018}
  {\path{doi:10.1016/j.cjph.2019.05.018}}.

\bibitem{Singh:2019ktp}
K.~Singh, S.~K. Maurya, M.~K. Jasim, F.~Rahaman, {Minimally deformed
  anisotropic model of class one space-time by gravitational decoupling}, Eur.
  Phys. J. C79~(10) (2019) 851.
\newblock \href {http://dx.doi.org/10.1140/epjc/s10052-019-7377-0}
  {\path{doi:10.1140/epjc/s10052-019-7377-0}}.

\bibitem{Abellan:2020wjw}
G.~Abellán, V.~Torres, E.~Fuenmayor, E.~Contreras, {Regularity condition on
  the anisotropy induced by gravitational decoupling in the framework of MGD},
  Eur. Phys. J. C80~(2) (2020) 177.
\newblock \href {http://arxiv.org/abs/2001.08573} {\path{arXiv:2001.08573}},
  \href {http://dx.doi.org/10.1140/epjc/s10052-020-7749-5}
  {\path{doi:10.1140/epjc/s10052-020-7749-5}}.

\bibitem{Hawking:1974sw}
S.~W. Hawking, {Particle Creation by Black Holes}, Commun. Math. Phys. 43
  (1975) 199--220, [,167(1975)].
\newblock \href {http://dx.doi.org/10.1007/BF02345020, 10.1007/BF01608497}
  {\path{doi:10.1007/BF02345020, 10.1007/BF01608497}}.

\bibitem{Bekenstein:1972tm}
J.~D. Bekenstein, {Black holes and the second law}, Lett. Nuovo Cim. 4 (1972)
  737--740.
\newblock \href {http://dx.doi.org/10.1007/BF02757029}
  {\path{doi:10.1007/BF02757029}}.

\bibitem{Bekenstein:1973ur}
J.~D. Bekenstein, {Black holes and entropy}, Phys. Rev. D7 (1973) 2333--2346.
\newblock \href {http://dx.doi.org/10.1103/PhysRevD.7.2333}
  {\path{doi:10.1103/PhysRevD.7.2333}}.

\bibitem{Hawking:1974rv}
S.~W. Hawking, {Black hole explosions}, Nature 248 (1974) 30--31.
\newblock \href {http://dx.doi.org/10.1038/248030a0}
  {\path{doi:10.1038/248030a0}}.

\bibitem{Wald:1984rg}
R.~M. Wald, {General Relativity}, Chicago Univ. Pr., Chicago, USA, 1984.
\newblock \href {http://dx.doi.org/10.7208/chicago/9780226870373.001.0001}
  {\path{doi:10.7208/chicago/9780226870373.001.0001}}.

\bibitem{Kastor:2009wy}
D.~Kastor, S.~Ray, J.~Traschen, {Enthalpy and the Mechanics of AdS Black
  Holes}, Class. Quant. Grav. 26 (2009) 195011.
\newblock \href {http://arxiv.org/abs/0904.2765} {\path{arXiv:0904.2765}},
  \href {http://dx.doi.org/10.1088/0264-9381/26/19/195011}
  {\path{doi:10.1088/0264-9381/26/19/195011}}.

\bibitem{Kothawala:2007em}
D.~Kothawala, S.~Sarkar, T.~Padmanabhan, {Einstein's equations as a
  thermodynamic identity: The Cases of stationary axisymmetric horizons and
  evolving spherically symmetric horizons}, Phys. Lett. B652 (2007) 338--342.
\newblock \href {http://arxiv.org/abs/gr-qc/0701002}
  {\path{arXiv:gr-qc/0701002}}, \href
  {http://dx.doi.org/10.1016/j.physletb.2007.07.021}
  {\path{doi:10.1016/j.physletb.2007.07.021}}.

\bibitem{Aros:2019auf}
M.~Estrada, R.~Aros, {Regular black holes with $\Lambda>0$ and its evolution in
  Lovelock gravity}, Eur. Phys. J. C79~(10) (2019) 810.
\newblock \href {http://arxiv.org/abs/1906.01152} {\path{arXiv:1906.01152}},
  \href {http://dx.doi.org/10.1140/epjc/s10052-019-7316-0}
  {\path{doi:10.1140/epjc/s10052-019-7316-0}}.

\bibitem{Dymnikova:2010zz}
I.~Dymnikova, M.~Korpusik, {Regular black hole remnants in de Sitter space},
  Phys. Lett. B685 (2010) 12--18.
\newblock \href {http://dx.doi.org/10.1016/j.physletb.2010.01.044}
  {\path{doi:10.1016/j.physletb.2010.01.044}}.

\bibitem{Kothawala:2009kc}
D.~Kothawala, T.~Padmanabhan, {Thermodynamic structure of Lanczos-Lovelock
  field equations from near-horizon symmetries}, Phys. Rev. D79 (2009) 104020.
\newblock \href {http://arxiv.org/abs/0904.0215} {\path{arXiv:0904.0215}},
  \href {http://dx.doi.org/10.1103/PhysRevD.79.104020}
  {\path{doi:10.1103/PhysRevD.79.104020}}.

\bibitem{Sheykhi:2010zz}
A.~Sheykhi, {Thermodynamics of apparent horizon and modified Friedmann
  equations}, Eur. Phys. J. C69 (2010) 265--269.
\newblock \href {http://arxiv.org/abs/1012.0383} {\path{arXiv:1012.0383}},
  \href {http://dx.doi.org/10.1140/epjc/s10052-010-1372-9}
  {\path{doi:10.1140/epjc/s10052-010-1372-9}}.

\bibitem{Padmanabhan:2002sha}
T.~Padmanabhan, {Classical and quantum thermodynamics of horizons in
  spherically symmetric space-times}, Class. Quant. Grav. 19 (2002) 5387--5408.
\newblock \href {http://arxiv.org/abs/gr-qc/0204019}
  {\path{arXiv:gr-qc/0204019}}, \href
  {http://dx.doi.org/10.1088/0264-9381/19/21/306}
  {\path{doi:10.1088/0264-9381/19/21/306}}.

\bibitem{Cai:2007bh}
R.-G. Cai, {Thermodynamics of apparent horizon in brane world scenarios}, Prog.
  Theor. Phys. Suppl. 172 (2008) 100--109.
\newblock \href {http://arxiv.org/abs/0712.2142} {\path{arXiv:0712.2142}},
  \href {http://dx.doi.org/10.1143/PTPS.172.100}
  {\path{doi:10.1143/PTPS.172.100}}.

\bibitem{Hayward:2005gi}
S.~A. Hayward, {Formation and evaporation of regular black holes}, Phys. Rev.
  Lett. 96 (2006) 031103.
\newblock \href {http://arxiv.org/abs/gr-qc/0506126}
  {\path{arXiv:gr-qc/0506126}}, \href
  {http://dx.doi.org/10.1103/PhysRevLett.96.031103}
  {\path{doi:10.1103/PhysRevLett.96.031103}}.

\bibitem{Dymnikova:1992ux}
I.~Dymnikova, {Vacuum nonsingular black hole}, Gen. Rel. Grav. 24 (1992)
  235--242.
\newblock \href {http://dx.doi.org/10.1007/BF00760226}
  {\path{doi:10.1007/BF00760226}}.

\bibitem{Aros:1999kt}
R.~Aros, M.~Contreras, R.~Olea, R.~Troncoso, J.~Zanelli, {Conserved charges for
  even dimensional asymptotically AdS gravity theories}, Phys. Rev. D62 (2000)
  044002.
\newblock \href {http://arxiv.org/abs/hep-th/9912045}
  {\path{arXiv:hep-th/9912045}}, \href
  {http://dx.doi.org/10.1103/PhysRevD.62.044002}
  {\path{doi:10.1103/PhysRevD.62.044002}}.

\bibitem{Kastor:2008xb}
D.~Kastor, {Komar Integrals in Higher (and Lower) Derivative Gravity}, Class.
  Quant. Grav. 25 (2008) 175007.
\newblock \href {http://arxiv.org/abs/0804.1832} {\path{arXiv:0804.1832}},
  \href {http://dx.doi.org/10.1088/0264-9381/25/17/175007}
  {\path{doi:10.1088/0264-9381/25/17/175007}}.

\bibitem{Chirco:2010rq}
G.~Chirco, S.~Liberati, T.~P. Sotiriou, {Gedanken experiments on nearly
  extremal black holes and the Third Law}, Phys. Rev. D82 (2010) 104015.
\newblock \href {http://arxiv.org/abs/1006.3655} {\path{arXiv:1006.3655}},
  \href {http://dx.doi.org/10.1103/PhysRevD.82.104015}
  {\path{doi:10.1103/PhysRevD.82.104015}}.

\end{thebibliography}

\end{document}